\documentclass[a4paper,english,american,aps,amsfonts,amsmath,prd,tightenlines, nofootinbib]{revtex4}
\usepackage[T1]{fontenc}
\usepackage[latin1]{inputenc}
\usepackage{amsmath}
\usepackage{amssymb}

\makeatletter
\usepackage{epsf}

\def\be{\begin{equation}}
\def\ee{\end{equation}}
\def\beq{\begin{equation}}
\def\eeq{\end{equation}}
\def\bea{\begin{eqnarray}}
\def\eea{\end{eqnarray}}
\def\bml{\begin{subequations}}
\def\blea{\bml\begin{eqnarray}}
\def\elea{\end{eqnarray}\end{subequations}}

\usepackage{babel}
\makeatother
\begin{document}

\title{Gravitational waves in vector inflation}

\author{Alexey Golovnev}

\author{Viatcheslav Mukhanov}

\author{Vitaly Vanchurin}

\affiliation{Arnold-Sommerfeld-Center for Theoretical Physics, Department f\"{u}r
Physik, Ludwig-Maximilians-Universit\"{a}t M\"{u}nchen, Theresienstr.
37, D-80333, Munich, Germany}

\begin{abstract}
We discuss the gravitational waves (GW) in the context of vector inflation.
We derive the action for tensor perturbations and find that tachyonic
instabilities are present in most (but not all) of the inflationary
models with large fields. In contrast, the stability of the small
field inflation ($A_{\mu}A^{\mu}\ll\frac{1}{N}$) is ensured by the
usual slow-roll conditions, where $N$ is the total number of fields.
For example, the Coleman-Weinberg potential and the power-law inflation
are always stable in the small fields limit with an approximately
flat spectrum of GW. We also provide some examples which lead to a
rapid decay of GW and predict the absence of tensor modes in the CMB.
\end{abstract}
\maketitle

\section{Introduction}

In the earlier paper \cite{GMV} we proposed a new model of inflation
in which the quasi de Sitter expansion is driven by vector fields.
Isotropy was achieved by employing a triad of mutually orthogonal
vector fields or by a large number of randomly oriented fields. The
problem of slow-roll was solved by non-minimal coupling of the vector
fields to gravity. (Other cosmological models with vector fields have
also been recently proposed \cite{BNO,DK,DLR,KKSY,KM,MB}).

The theory of Ref. \cite{GMV} is defined by the following action\begin{align}
S & =\int\sqrt{-g}\left[-\frac{R}{2}\left(1+\sum_{a=1}^{N}\frac{1}{6}I_{(a)}\right)-\frac{1}{4}\sum_{a=1}^{N}F_{\mu\nu}^{(a)}F_{(a)}^{\mu\nu}-\sum_{a=1}^{N}V\left(I_{(a)}\right)\right]dx^{4}.\label{eq:action1}\end{align}
where\begin{align*}
I_{(a)} & \equiv{-A}_{\mu}^{(a)}A_{(a)}^{\mu}\\
F_{\mu\nu}^{(a)} & \equiv\nabla_{\mu}A_{\nu}^{(a)}-\nabla_{\nu}A_{\mu}^{(a)}\end{align*}
and summation over repeated space-time indices is assumed. In the
spatially flat Friedmann universe with conformal metric\[
ds^{2}=a^{2}\left(\eta\right)\left(d\eta^{2}-\delta_{ik}dx_{i}dx_{k}\right)\]
the evolution of homogeneous background fields is given by\begin{align}
A_{0} & =0\nonumber \\
{B''}_{i}+2{\cal H}{B'}_{i} & +2\frac{dV\left(I\right)}{dI}a^{2}B_{i}=0\label{eq:field}\end{align}
where by prime we denote the differentiation with respect to conformal
time $\eta$ and $I=B_{i}B_{i}\equiv B^{2}$, $B_{i}\equiv\frac{A_{i}}{a}=-aA^{i}$,
${\cal H}\equiv\frac{a'}{a}$. Equation (\ref{eq:field}) is equivalent
to

\[
{A''}_{i}+\left(2a^{2}\frac{dV}{dI}-\frac{a''}{a}\right)A_{i}=0.\]
The Einstein equations reduce to \begin{equation}
{3{\cal H}}^{2}=8\pi N\left(V\left(B\right)a^{2}+\frac{1}{2}B'^{2}\right),\label{eq:einstein}\end{equation}
\begin{equation}
2{\cal H}^{\prime}+{\cal H}^{2}=8\pi N\left(V(B^{2})a^{2}-\frac{1}{2}B^{\prime2}\right)\label{eq:einstein2}\end{equation}
and for the usual mass term we have $V=\frac{m^{2}B^{2}}{2}=-\frac{m^{2}A_{\mu}A^{\mu}}{2}$
and $\frac{dV}{dI}=\frac{m^{2}}{2}$.

The paper is organized as follows. In the next section we derive the
action for gravitational waves. In the third section we quantize the
tensor perturbations and solve the corresponding equations of motion.
The consequences of our results for different models of vector inflation
are analyzed in the forth section and the main conclusions are summarized
in the final section.

\section{Action for gravitational waves}

We consider the transverse and traceless metric perturbations $h_{ik}$
on a spatially flat Friedmann background \[
ds^{2}={a\left(\eta\right)}^{2}\left(d\eta^{2}-\left(\delta_{ik}-h_{ik}\right)dx_{i}dx_{k}\right)\]
where $h_{\phantom{i}i}^{i}=0$ and $h_{\phantom{i}j,i}^{i}=0.$ As
the matter content is of vectorial nature, these are the only tensor
perturbations of the theory. The corresponding equation of motion
can be obtained as a tensor part of the spatial components of Einstein
equations. One should note that scalar, vector and tensor perturbations
can couple to the background vectors \cite{Armendaris}. Some of these
couplings vanish due to the background rotational symmetry. For example,
$\sum_{a=1}^{N}h_{ik}A_{i}^{(a)}A_{k}^{(a)}=0$ because it is proportional
to the trace of $h$, which is a consequence of the isotropy condition
$\sum_{a=1}^{N}A_{i}^{(a)}A_{k}^{(a)}\propto\delta_{ik}$ and follows
from the fact that the trace is the only linear rotational invariant
of a matrix. Nevertheless, terms of the form $A_{i}{\delta A}_{k}$
are present in $T_{i}^{k}$ and in general do not decouple from GW.
However, if we consider random vector fields with random fluctuations,
these terms would be statistically suppressed.%
\footnote{Terms of the form $A_{i}{\delta A}_{k}$are exactly canceled for perturbations
which correspond to rotations of a vector triad configuration as a
whole.%
} This allows us to consider the evolution of GW separately from the
vector and scalar perturbations.

From now on, in this article, we consider only the tensor perturbations.
To obtain the action for these perturbations we must expand Eq. (\ref{eq:action1})
to the second order in $h_{ik}$. The spatial part of the metric is
given by $g_{ij}=-\delta_{ij}+h_{ij}$ and $g^{ij}=-\delta_{ij}-h_{ij}-h_{ik}h_{kj}$
up to the second order. It implies \[
\sqrt{-g}=a^{4}\left(1-\frac{1}{4}h_{ij}^{2}\right),\]

\[
R=\frac{1}{a^{2}}\left(-6\frac{a^{\prime\prime}}{a}+h_{ij}h_{ij}^{\prime\prime}+\frac{3}{4}h_{ij}^{\prime2}-h_{ij}\bigtriangleup h_{ij}-\frac{3}{4}h_{ij,k}^{2}+\frac{1}{2}\left(h_{ij}h_{ik,j}\right)_{,k}+3h_{ij}h_{ij}^{\prime}H\right)\]
and

\begin{align}
S_{gw} & \approx\frac{1}{8}\int a^{2}\left[\left(\frac{1}{8\pi}+\frac{NB^{2}}{6}\right)\left({h'}_{ik}^{2}-h_{ik,j}^{2}-m_{g}^{2}h_{ik}^{2}\right)\right]dx^{3}d\eta\label{eq:action-gw}\end{align}
(where we used the background equations, averaging and integrations
by parts). The graviton mass squared is given by \begin{align}
m_{g}^{2} & \equiv-16\pi N\frac{\left(\frac{a''}{a^{3}}-2V_{,I}-\frac{4}{5}B^{2}V_{,II}\right)a^{2}B^{2}+\left(B'+B{\cal H}\right)^{2}}{3+4\pi NB^{2}}\label{eq:gravition_mass}\end{align}
where $V_{,I}\equiv\frac{dV}{dI}$, $V_{,II}\equiv\frac{d^{2}V}{dI^{2}}$.
Unusual mass of the graviton comes from the terms $A_{\mu}A_{\nu}g^{\mu\nu}$
and $F_{\mu\nu}F_{\alpha\beta}g^{\mu\alpha}g^{\nu\beta}$. It is proportional
to $B^{2}H^{2}$ which remains approximately constant during slow-roll
inflation. 

The corresponding equation of motion (which could also be derived
form the linearized Einstein equation) is

\begin{align}
{h''}_{ik}+2\left({\cal H}+\frac{4\pi NBB^{\prime}}{3+4\pi NB^{2}}\right){h'}_{ik}-\bigtriangleup h_{ik} & =-m_{g}^{2}h_{ik}.\label{eq:waves}\end{align}
During the slow-roll inflation $\ddot{B}\ll H\dot{B}$ implies $B^{\prime\prime}\approx{\cal H}B^{\prime}$
and we obtain \begin{equation}
m_{g}^{2}\approx16\pi N\frac{\left(-8\pi NV+\frac{10}{3}V_{,I}+\frac{4}{5}B^{2}V_{,II}\right)a^{2}B^{2}}{3+4\pi NB^{2}}\label{eq:gravition_mass_slow_roll}\end{equation}
which can be further reduced to\[
m_{g}^{2}\approx16\pi m^{2}a^{2}NB^{2}\left(\frac{5-12\pi NB^{2}}{9+12\pi NB^{2}}\right)\]
for the chaotic potential $V\left(B^{2}\right)=\frac{1}{2}m^{2}B^{2}$.
In the original paper \cite{GMV} we have shown that $B\gtrsim\frac{1}{\sqrt{N}}$
for this potential, which means that the evolution can be unstable
due to the tachyonic mass of the graviton $m_{g}^{2}<0$. We will
come back to the stability issue in the forth section after the theory
is properly quantized.

\section{Creation and evolution of perturbations}

We follow the procedure outlined in Ref. \cite{book} to quantize
the action for GW (\ref{eq:action-gw}). The first step is to expand
the tensor perturbations into Fourier modes\begin{equation}
h_{ij}\left(\mathbf{x},\eta\right)=\int h_{\mathbf{k}}\left(\eta\right)e_{ij}\left(\mathbf{k}\right)e^{i\mathbf{k}\mathbf{x}}\frac{d^{3}k}{\left(2\pi\right)^{3/2}},\label{eq:Fourier}\end{equation}
where $e_{ij}\left(\mathbf{k}\right)$ is the polarization tensor.
The result is substituted in Eq. (\ref{eq:action-gw}) to obtain\[
S_{gw}\approx\int\frac{1}{8}a^{2}e_{ij}^{2}\left[\left(\frac{1}{8\pi}+\frac{NB^{2}}{6}\right)\left({h'}_{\mathbf{k}}{h'}_{-\mathbf{k}}-\left(k^{2}+m_{g}^{2}\right)h_{\mathbf{k}}h_{-\mathbf{k}}\right)\right]dk^{3}d\eta.\]
It is convenient to introduce a new variable\begin{equation}
v_{\mathbf{k}}=\frac{1}{2}ah_{\mathbf{k}}\sqrt{e_{ij}^{2}\left(\frac{1}{8\pi}+\frac{NB^{2}}{6}\right)}\label{eq:v_k_defined}\end{equation}
and rewrite the action as \[
S_{gw}\approx\frac{1}{2}\int\left[{v'}_{\mathbf{k}}{v'}_{-\mathbf{k}}-\omega_{k}^{2}\left(\eta\right)v_{\mathbf{k}}v_{-\mathbf{k}}\right]dk^{3}d\eta\]
Assuming the slow-roll regime we convert the equation of motion (\ref{eq:field})
into $B^{\prime}=-\frac{{2a}^{2}V_{,I}B}{3{\cal H}}$ while the Friedman
equations (\ref{eq:einstein}) and (\ref{eq:einstein2}) yield $\frac{a^{\prime\prime}}{a}=\frac{16\pi}{3}Na^{2}V$
and one can easily deduce that \begin{align*}
\omega_{k}^{2}\left(\eta\right) & \equiv k^{2}-\frac{a''\left(\eta\right)}{a\left(\eta\right)}\beta\end{align*}
with\begin{align}
\beta & \equiv1+\frac{24\pi NB^{2}-\left(\frac{23}{2}\frac{V_{,I}}{V}B^{2}+\frac{12}{5}\frac{V_{,II}}{V}B^{4}\right)}{4\pi NB^{2}+3}.\label{eq:beta}\end{align}
The corresponding equation of motion is \begin{align}
{v''}_{\mathbf{k}}+\omega_{k}^{2}\left(\eta\right)v_{\mathbf{k}} & =0.\label{eq:v_k_equation}\end{align}
After the usual quantization procedure we obtain the standard power
spectrum of the created waves\begin{equation}
\delta_{h}^{2}\left(k,\eta\right)=\frac{{8\left|v_{\mathbf{k}}\right|}^{2}k^{3}}{\pi a^{2}}.\label{eq:spectrum}\end{equation}

To solve the Eq. (\ref{eq:v_k_equation}) during the slow-roll inflation
we can use an approximation, where $\beta\approx const$ and $\frac{a''}{a}\simeq\frac{2}{\eta^{2}}$.
The general solution is given by \[
v_{\mathbf{k}}\left(\eta\right)=\sqrt{\eta}\left(C_{1}J_{\frac{\sqrt{1+8\beta}}{2}}\left(k\eta\right)+C_{2}Y_{\frac{\sqrt{1+8\beta}}{2}}\left(k\eta\right)\right)\]
where $J$ and $Y$ are Bessel functions of the first and second kind
respectively. For the short-wavelength modes with $k^{2}\gg\left|\frac{a''}{a}\beta\right|$
we obtain the usual result\[
v_{\mathbf{k}}\left(\eta\right)\simeq\frac{1}{\sqrt{k}}\, e^{\pm ik(\eta-\eta_{i})}\]
and for the long-wavelength perturbations with $k^{2}\ll\left|\frac{a''}{a}\beta\right|$
the solution becomes \[
v_{\mathbf{k}}\left(\eta\right)\simeq C_{1}\,\eta^{\frac{1-\sqrt{1+8\beta}}{2}}+C_{2}\,\eta^{\frac{1+\sqrt{1+8\beta}}{2}}.\]
In the limit of $\beta\rightarrow1$, the evolution of tensor perturbation
is identical to the scalar field inflation. The non-decaying super-horizon
modes of GW ($h_{\mathbf{k}}\propto\frac{v_{\mathbf{k}}}{a}$) are
frozen, and from Eq. (\ref{eq:spectrum}) we get an approximately
flat power spectrum with a slightly red tilt. However in general the
tensor perturbations could grow with time if $\beta>1$, and decay
if $\beta<1$ which would lead to somewhat different predictions.

\section{Gravitational waves in different models}

\selectlanguage{english}
Now we are in a position to analyze the behavior of GW in different
inflationary scenarios. The key ingredient of our discussion is the
expression (\ref{eq:beta}) for $\beta$, which can be simplified
in the two limiting cases corresponding to the large and small field
approximations. 

If the inflationary evolution takes place at large values of the field
($B\gg\frac{1}{\sqrt{N}}$), then \foreignlanguage{american}{\begin{align}
\beta & \approx7-\frac{1}{4\pi N}\left(\frac{23}{2}\frac{V_{,I}}{V}+\frac{12}{5}B^{2}\frac{V_{,II}}{V}\right).\label{eq:beta_large}\end{align}
Such models would generically predict $\beta\sim7$ leading to} very
large instabilities of GW, incompatible with observations. Nevertheless,
for some potentials the second term on the right-hand side of Eq.
(\ref{eq:beta_large}) can be large enough to reduce $\beta$ towards
the observationally allowed range ($\beta\lesssim1$). \foreignlanguage{american}{Unfortunately,
as we shall see, it is rather hard to obtain a working model of vector
inflation with large fields. }

\selectlanguage{american}
A much more promising class of models describes the evolution at small
values of the inflation field. From Eq. \foreignlanguage{english}{(\ref{eq:beta}),
in the limit $B\ll\frac{1}{\sqrt{N}}$,} we obtain \foreignlanguage{english}{}\begin{equation}
\beta\approx1-\left(\frac{23}{6}\frac{V_{,I}}{V}B^{2}+\frac{4}{5}\frac{V_{,II}}{V}B^{4}\right).\label{eq:beta_small}\end{equation}
Clearly, all of the models with $\frac{23}{6}\frac{V_{,I}}{V}B^{2}+\frac{4}{5}\frac{V_{,II}}{V}B^{4}\ll1$
would predict a stable evolution with nearly flat spectrum of tensor
perturbations, similarly to the standard scalar field inflation. In
fact, the usual slow-roll conditions \begin{align*}
\frac{V_{,B}}{V} & =2\frac{V_{,I}}{V}B\ll1\\
\frac{V_{,BB}}{V} & =2\frac{V_{,I}}{V}+4\frac{V_{,II}}{V}B^{2}\ll1\end{align*}
automatically imply \begin{align*}
\frac{23}{6}\frac{V_{,I}}{V}B^{2}\ll\frac{23}{6}B & <1\\
\frac{4}{5}\frac{V_{,II}}{V}B^{4}\ll & 1\end{align*}
in the limit of small fields $B\ll\frac{1}{\sqrt{N}}$ and for a large
number of fields $N\gtrsim15$. However, for a relatively small number
of fields $N\lesssim15$ one should also keep track of the parameter
$\frac{23}{6}\frac{V_{,I}}{V}B^{2}$, which could be of order one. 

\selectlanguage{english}

\subsection{Chaotic potential}

Consider a model of chaotic vector inflation ($V=\frac{m^{2}B^{2}}{2}$)
proposed in Ref. \cite{GMV}. It follows from Eq. (\ref{eq:beta_large})
that during inflation $\frac{1}{\sqrt{2\pi N}}<B<\frac{1}{N^{1/4}}$
and the parameter $\beta$ stays in between $\beta_{i}\approx7$ and
$\beta_{f}\approx5$. At the beginning of inflation the contribution
of the terms $\frac{V_{,I}}{V}B^{2}$ and $\frac{V_{,II}}{V}B^{4}$
is suppressed by a factor of $\frac{1}{\sqrt{N}}$, and $\beta$ changes
very slowly down from the value of $7$. (An additional problem of
the chaotic vector inflation in the spatially curved universes was
discussed in Ref. \cite{Chiba}.)

The instability can be fixed by going to higher powers of $B^{2}$.
For a potential $V=V_{0}B^{2n}$ with a very large $n>300$ we can
(in principle) have a large number of e-folds ${\cal N}=\frac{2\pi N(B_{i}^{2}-B_{f}^{2})}{n}>60$
with relatively small (or negative) values of $\beta\lesssim7-\frac{6}{5}\frac{n}{{\cal N}}$,
predicting the absence of tensor modes in the CMB. However, such models
seem to be rather fine-tuned and we shall not discuss them any further.
Instead we will turn our attention to the models with small field
inflation which naturally predict a stable evolution of GW.

\selectlanguage{american}

\subsection{Power-law inflation}

\selectlanguage{english}
An interesting example describes the power-law inflation with potential
$V=V_{0}\exp\left(\alpha\sqrt{A_{\mu}A^{\mu}}\right)=V_{0}\exp\left(\alpha\sqrt{B^{2}}\right)$,
where we need $\alpha\leq2\sqrt{6\pi N}$ for $p\leq-\frac{\epsilon}{3}$.
(One could worry that this potential is not regular at $A_{\mu}=0$,
but in either case one has to modify the potential at small $A_{\mu}$
for the inflation to end.) For this model, only if the inflation takes
place at small values of $B\ll\frac{1}{\sqrt{N}}$ then $\beta\sim1$
and the evolution is stable.

\subsection{Symmetry-breaking potential}

For the symmetry-breaking potential $V=\lambda\left(B^{2}-B_{0}^{2}\right)^{2}$
the evolution starts at some small value of $B_{i}$ with $\beta_{i}$
close to $1$ and ends when $p\sim-\frac{\epsilon}{3}$ where $p=\frac{N}{2}\left(\dot{B}^{2}-2V\right)$
and $\epsilon=\frac{N}{2}\left(2V+\dot{B}^{2}\right)$. The number
of e-folds in the slow-roll and small-fields approximation is given
by \[
{\cal N}\approx-\pi NB_{0}^{2}\ln\left(\frac{8\pi}{3}NB_{0}^{2}\lambda\right).\]
 where the initial value of the field was set by quantum fluctuations
to $B_{i}\sim H\approx B_{0}^{2}\sqrt{\frac{8\pi}{3}N\lambda}$. Apparently
it is hard to obtain a long period of inflation with small fields
because of the logarithmic dependence on $\lambda$. For $NB_{0}^{2}\sim0.1$
we must have $\lambda\lesssim10^{-60}$ in order to get at least $60$
e-folds of inflation.

\subsection{Coleman-Weinberg potential}

A better illustration of the vector inflation is given by the Coleman-Weinberg
potential $V=\lambda\left(B^{4}\ln\frac{B^{2}}{B_{0}^{2}}-\frac{1}{2}B^{4}+\frac{1}{2}B_{0}^{4}\right)$,
where a large number of e-folds can naturally occur at small values
of the field ($B<B_{0}\ll\frac{1}{\sqrt{N}}$). Indeed, the number
of e-folds is

\[
{\cal N}=-2\pi N\int_{B_{i}^{2}}^{B_{f}^{2}}\frac{V}{V_{,I}B^{2}}dB^{2}\approx\frac{\pi{NB}_{0}^{2}}{2}\frac{\frac{B_{0}^{2}}{B_{i}^{2}}}{\ln\frac{B_{0}^{2}}{B_{i}^{2}}}\]
where we assume $B_{i}\ll B_{0}$. In fact, the initial value $B_{i}$
is determined by quantum fluctuations $B_{i}\sim H$ near the maximum
$V\approx\frac{\lambda}{2}B_{0}^{4}$ which gives $B_{i}\approx B_{0}^{2}\sqrt{\frac{4\pi}{3}N\lambda}$
implying $\frac{B_{0}^{2}}{B_{i}^{2}}=\frac{3}{4\pi\lambda NB_{0}^{2}}$.
Neglecting the logarithmic dependence, the number of e-folds 

\[
{\cal N}\approx\frac{3}{8\lambda}\cdot\left(\ln\frac{3}{4\pi\lambda NB_{0}^{2}}\right)^{-1}\]
scales as $\lambda^{-1}$ (in contrast to the logarithmic dependence
for the symmetry breaking potential). For small values of $\lambda$
the model can produce an arbitrary large number of e-folds of stable
inflation with an approximately flat power spectrum of GW.

\subsection{Exponential potential}

\selectlanguage{american}
Another example is provided by an exponential potential $V=\lambda\exp\left(-\alpha B^{2}\right)$.
The small field assumption ($B\ll\frac{1}{\sqrt{N}}$) is not generally
valid all the way until the end of inflation $B_{f}=\frac{\sqrt{6\pi N}}{\alpha}$,
and one has to work in the limit of large fields. The inflation in
such models is typically very long, and a large number of e-folds
take place at negative values of $\beta$. For example, if we take
$\alpha=\frac{N}{50}$ then the exit from inflation occurs at $B_{f}\approx\frac{217}{\sqrt{N}}$
and much more than the last $60$ e-folds evolve at negative values
of $\beta$. Such models would generically predict the absence of
tensor modes in the CMB.

\section{Conclusions}

The recently proposed model of vector inflation \cite{GMV} was shown
to be very similar to the scalar field inflation at the background
level. Despite of the apparent similarities a number of striking differences
appear already at the first order in perturbation theory. The most
important difficulty is the coupling of different modes (scalar, vector
and tensor) even in the linear order. One can overcome the problem
for the tensor perturbations in the limit of a large number of random
fields with random fluctuations, when the coupling terms are statistically
suppressed. 

In this article we concentrated only on the behavior of gravitational
waves on the homogeneous and isotropic background. \foreignlanguage{english}{Our
analysis has shown that the behavior of tensor perturbations in vector
inflation depends crucially on the form of the inflationary potential.
The large field inflationary models generically lead to the tachyonic
instabilities of GW (e.g. chaotic potential) and must be carefully
analyzed. The stability of the small fields inflationary models is
guaranteed whenever the slow-roll conditions are satisfied. In addition,
the models of vector inflation can lead to an arbitrary tilt in the
power spectrum as well as to a complete suppression of the tensor
modes in the CMB. }

\begin{acknowledgments}
The authors are grateful to Misao Sasaki, Alexander Vikman and Sergei
Winitzki for helpful discussions. This work was supported in part
by TRR 33 {}``The Dark Universe'' and the Cluster of Excellence
EXC 153 {}``Origin and Structure of the Universe''. 
\end{acknowledgments}

\end{document}